\documentclass[pra, a4paper, aps, twocolumn, notitlepage, superscriptaddress, floatfix, longbibliography]{revtex4-2}
\usepackage[breaklinks]{hyperref}
\usepackage{etoolbox}
\appto\UrlBreaks{\do\-}
\usepackage{amsmath, amssymb, amsthm}
\usepackage{mathtools}
\usepackage{color}
\usepackage{braket}
\usepackage{bbm}
\usepackage{nicefrac}
\usepackage{booktabs}
\usepackage[whole]{bxcjkjatype}
\allowdisplaybreaks[4]

\theoremstyle{definition}

\theoremstyle{definition}

\theoremstyle{remark}

\begin{document}

\title{Constant-time one-shot testing of large-scale graph states}

\author{Hayata Yamasaki}
\email{hayata.yamasaki@gmail.com}
\affiliation{Institute for Quantum Optics and Quantum Information --- IQOQI Vienna, Austrian Academy of Sciences, Boltzmanngasse 3, 1090 Vienna, Austria}
\affiliation{Atominstitut,  Technische  Universit{\"a}t  Wien, Stadionallee 2, 1020  Vienna,  Austria}
\author{Sathyawageeswar Subramanian}
\email{sathya.subramanian@warwick.ac.uk}
\affiliation{Department of Computer Science, University of Warwick, Coventry CV4 7EZ, United Kingdom}

\begin{abstract}
Fault-tolerant measurement-based quantum computation (MBQC) with recent progress on quantum technologies leads to a promising scalable platform for realizing quantum computation, conducted by preparing a large-scale graph state over many qubits and performing single-qubit measurements on the state.
With fault-tolerant MBQC, even if the graph-state preparation suffers from errors occurring at an unknown physical error rate, we can suppress the effect of the errors.
Verifying graph states is vital to test whether we can conduct MBQC as desired even with such errors.
However, problematically, existing state-of-the-art protocols for graph-state verification by fidelity estimation have required measurements on many copies of the entire graph state and hence have been prohibitively costly in terms of the number of qubits and the runtime.
We here construct an efficient alternative framework for testing graph states for fault-tolerant MBQC based on the theory of property testing.
Our test protocol accepts with high probability when the physical error rate is small enough to make fault-tolerant MBQC feasible and rejects when the rate is above the threshold of fault-tolerant MBQC\@.
The novelty of our protocol is that we use only a single copy of the $N$-qubit graph state and single-qubit Pauli measurements only on a constant-sized subset of the qubits; thus, the protocol has a constant runtime independently of $N$.
Furthermore, we can immediately use the rest of the graph state for fault-tolerant MBQC if the protocol accepts.
These results achieve a significant advantage over prior art for graph-state verification in the number of qubits and the total runtime.
Consequently, our work offers a new route to a fast and practical framework for benchmarking large-scale quantum state preparation.
\end{abstract}

\maketitle

\textit{Introduction.}---
Measurement-based quantum computation (MBQC)~\cite{Raussendorf01,Raussendorf03,Jozsa06,Briegel2009} is a model of quantum information processing wherein computations are encoded into a sequence of adaptive measurements performed on subsystems of a fixed multipartite entangled state, a \textit{resource} state.
Except for its size (number of qubits), the resource state is independent of what to compute, and the computation is driven by changing the pattern of measurement bases conditioned on the outcomes of previous measurements.
Graph states~\cite{H13} are multiqubit entangled states widely used for MBQC\@. 
In particular, families of $N$-qubit graph states $\Ket{G_N}$ associated with graphs $G_N$ on $N$ vertices can be used for implementing arbitrary quantum gate sequences by a pattern of adaptive single-qubit measurements.
If a source of entangled states outputs the required large-scale graph state in high fidelity, then the MBQC protocol can achieve universal quantum computation with high success probability. 
Recent technological progress in quantum photonics achieves the preparation and measurement of entangled states over one million subsystems, leading to a promising scalable platform for realizing quantum computation via MBQC~\cite{Y4,A7,Larsen369,PhysRevApplied.16.034005,yamasaki2020polylogoverhead,PhysRevX.8.021054,Bourassa2021blueprintscalable,PhysRevLett.115.020502}.

\begin{figure}[tb]
    \centering
    \includegraphics[width=3.4in]{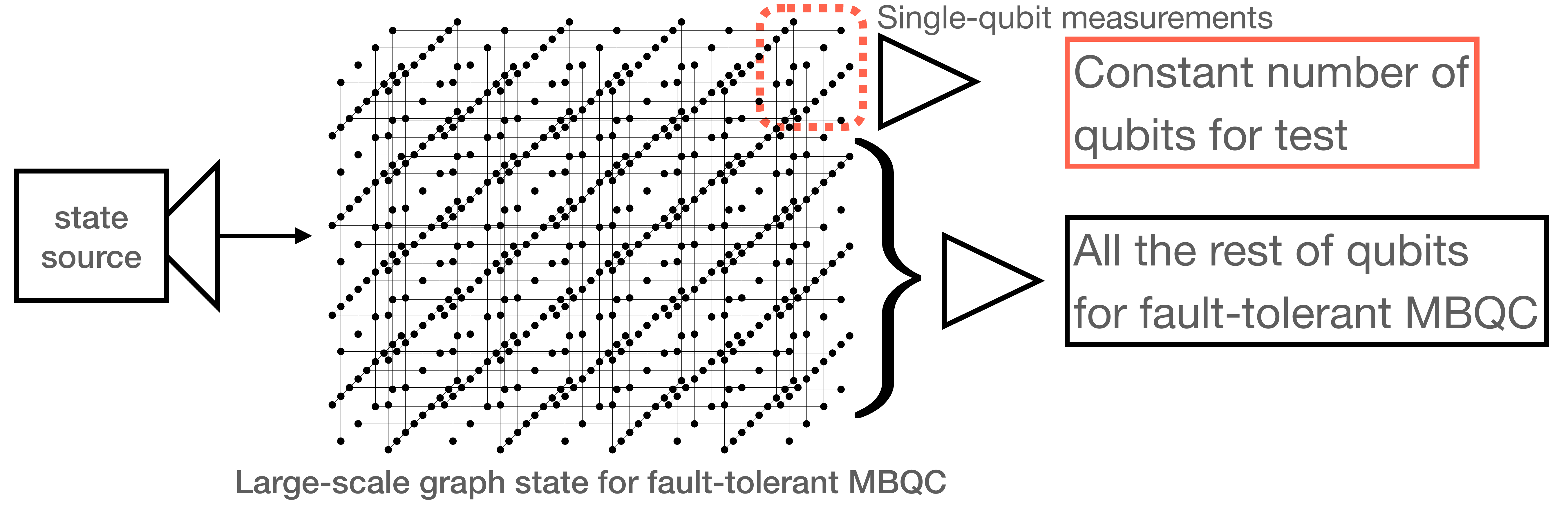}
    \caption{Framework for testing a large-scale graph state prepared by a source. As in a conventional setting of fault-tolerant quantum computation, the state from the source may suffer from IID Pauli errors. Our protocol for testing an $N$-qubit graph state performs single-qubit $X$- and $Z$-basis measurements on a constant number of qubits, independently of the size $N$ of the graph state. From the measurement outcomes, we calculate a success condition for the test within a constant time. If the physical error rate is above the threshold of fault-tolerant MBQC, the protocol rejects with high probability. On the other hand, if the physical error rate is small enough to perform fault-tolerant MBQC feasibly (i.e., bounded away from the threshold), the protocol accepts with high probability, and we can use the rest of this single copy of the graph state for fault-tolerant MBQC to implement universal quantum computation with arbitrarily suppressed logical error rate.}
    \label{fig:intro}
\end{figure}

In light of such technological progress, verification protocols for graph states have been attracting considerable attention~\cite{PhysRevLett.115.220502,PhysRevA.96.030301,PhysRevLett.120.170502,takeuchi2019resource,PhysRevApplied.12.054047,PhysRevLett.123.260504,PhysRevA.100.062335,markham2018simple,PhysRevResearch.2.043323}.
Since the graph state required for MBQC is on the large scale in terms of the number of qubits, estimating a full classical description of the state by quantum state tomography is prohibitively costly and practically infeasible.
On the other hand, for the feasibility of MBQC, it is unnecessary to have such a detailed exponential-sized description of the state $\rho$ emitted by the source; rather, it is sufficient to verify that $\rho$ is close to the required graph state $\Ket{G_N}$ in some distance measure, such as that based on fidelity.

Existing state-of-the-art protocols for fidelity estimation achieve this verification~\cite{PhysRevApplied.12.054047,PhysRevLett.115.220502,takeuchi2019resource,PhysRevLett.123.260504,PhysRevA.100.062335,PhysRevA.96.030301,markham2018simple,PhysRevLett.120.170502,PhysRevResearch.2.043323}. 
These protocols perform single-qubit measurements on $T$ copies of $\rho$ obtained from the source, and guarantee that if these measurement outcomes suffer from no error and satisfy a certain success condition, then with high probability the fidelity $F=\Bra{G_N}\rho\Ket{G_N}$ is at least $F=1-O(\nicefrac{1}{T})$, which achieves the optimal scaling in $\nicefrac{1}{T}$~\cite{PhysRevLett.120.170502}.
However, problematically, these protocols need a large number of copies of the \textit{entire} large-scale state, in addition to the one used in MBQC after the verification. 
For example, for a reasonable choice of parameters according to Fig.~2 of Ref.~\cite{PhysRevApplied.12.054047}, the required number of copies can be of the order of $T\approx 10^{3}$.
Moreover, these protocols may require that all the measurements made on the $T$ copies should simultaneously satisfy a success condition, which may be hard in practice due to the accumulation of unavoidable errors in measuring and manipulating quantum systems.
In principle, quantum error correction can be used to make all the $T$ copies satisfy the success condition with high probability~\cite{PhysRevA.96.030301}. However, to achieve this, the fault-tolerant protocol using quantum error correction would incur further overhead in preparing and manipulating encoded versions of each of the $T$ copies of $\rho$.
As a result, the existing protocols for the verification of graph states based on fidelity estimation may still be too challenging to perform in practice due to the required computational resources, i.e., the overhead in the sample complexity $T$, the number of qubits to be measured for the verification, and the total runtime.
A more efficient toolkit for verifying and benchmarking graph-state preparation is thus in high demand from both theoretical and practical perspectives, even allowing for reasonable modifications of the goal of \textit{estimating} the fidelity or other figures of merit.\\

\textit{This work.}---
Here we develop an alternative framework for efficiently testing the graph-state preparation, as shown in Fig.~\ref{fig:intro}.
In general, even a single-qubit bit- or phase-flip error in the state preparation may nullify the fidelity between $\rho$ prepared by the source and $\Ket{G_N}$ to be prepared.
But indeed, even if multiple independent and identically distributed (IID) errors occur over a constant fraction of the physical qubits, we may still be able to perform quantum error correction using a graph state from a special family of $3$D graphs~\cite{RAUSSENDORF20062242,PhysRevLett.98.190504,Raussendorf_2007,PhysRevLett.117.070501,nickerson2018measurement,Newman2020generatingfault}.
In a conventional setting of fault-tolerant quantum computation,
although the source in the ideal case is supposed to prepare a graph state required for fault-tolerant MBQC, the state may suffer from IID Pauli errors at some nonzero physical error rate $p$. If $p$ is below a certain threshold $p_\mathrm{th}$, the well-established protocol for fault-tolerant MBQC can arbitrarily suppress the logical error rate by quantum error correction~\cite{RAUSSENDORF20062242,PhysRevLett.98.190504,Raussendorf_2007,PhysRevLett.117.070501,nickerson2018measurement,Newman2020generatingfault}.

Considering this setting, we formulate a task that focuses on \textit{testing}, with high probability, whether the physical error rate is low enough to suppress the logical error rate feasibly or not even below the threshold. Note that this stands in contrast to verifying the input state's fidelity with the target graph state by fidelity \textit{estimation}.
We then construct a protocol that accomplishes this task with a drastic improvement in resource requirement compared to all known verification protocols based on fidelity estimation.
In particular, our protocol is a \textit{one-shot} protocol that uses only a \textit{single copy} of the input state ($T=1$).
Furthermore, for bounded-degree periodic graphs, i.e., all the known graph states for fault-tolerant MBQC~\cite{RAUSSENDORF20062242,PhysRevLett.98.190504,Raussendorf_2007,PhysRevLett.117.070501,nickerson2018measurement,Newman2020generatingfault}, we can perform our protocol only by non-adaptive single-qubit Pauli measurements on a constant number of qubits. As a bonus, if the measurement outcomes satisfy a simple success condition, the rest of qubits of the graph state can immediately be used for achieving fault-tolerant MBQC\@.  
Importantly, we can compute and check the success condition within a constant time independently of the total number $N$ of qubits of the entire graph state $\Ket{G_N}$ to be prepared.

Consequently, by virtue of having only a constant overhead in samples, qubits, and runtime, our results lead to a practically feasible framework for testing arbitrarily large-scale graph states for fault-tolerant MBQC\@.
It is also worth remarking that the testing framework, and the key ideas used in our protocol, are designed to be extendable to other more general classes of quantum states beyond graph states.\\

\textit{Connection to Property Testing.}---
Fast methods for solving relaxations of decision problems by only locally accessing a small fraction of the input fall under the broad purview of property-testing algorithms~\cite{DBLP:conf/propertytesting/2010}.
More specifically, the testing problem in this work is formulated as a gapped promise problem in property testing, wherein algorithms are required to be able to decide with high probability whether the input scores above a threshold $\alpha$ or below a threshold $\beta$ on some function of interest --- for example, the algorithm may want to decide if a given probability distribution on $n$ items has Shannon entropy larger than $\alpha\log_2 n$, or smaller than $\beta\log_2 n$, by only looking at few samples out of the $n$ items~\cite{10.1145/509907.510005,gur2021sublinear}.
Promise gap here refers to the situation that we are free to output a random decision when the input falls into the region of no interest in between the two thresholds.
Of crucial importance to algorithm design and complexity is the gap $\alpha-\beta$; intuitively, larger gaps imply easier relaxations of the underlying decision problem.

Our framework for testing graph states is formulated in this spirit of property testing: our protocol measures only a constant number of qubits of the arbitrarily large input state in order to decide the property of having a low or high physical error rate. The task we address recognizes that, to verify the goodness of an input state, it may not be strictly necessary in fault-tolerant MBQC to estimate a distance metric. Rather, when the information about the underlying error model is available, it is sufficient to directly test whether the physical error rate is small enough --- in particular, low physical error rates will in general automatically ensure high fidelity on logical qubits of quantum error-correcting codes in fault-tolerant MBQC\@.
We here construct our framework based on the fact that this physical error rate is a granular and local property, as opposed to distance metrics that are global properties of the entire input state.

Our results thus establish a hitherto unexplored connection between the complexity theoretic study of property-testing algorithms in theoretical computer science and the manifestly practical task of benchmarking the preparation of large-scale quantum states in physical experiments.\\

%%%%%%%%%%%%%%%%%%%%%%
\begin{figure}
    \centering
    \includegraphics[width=3.4in]{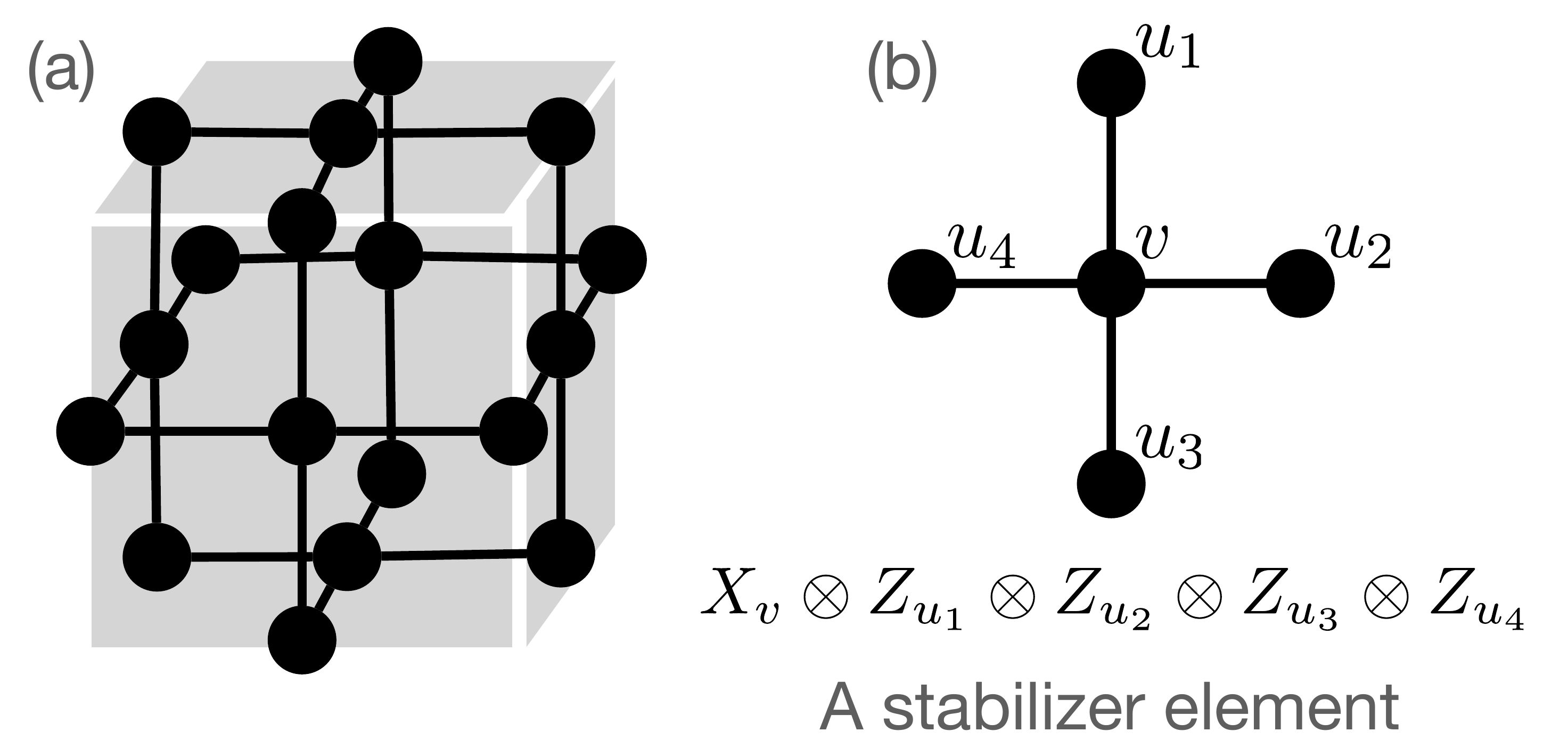}
    \caption{(a) The unit cell of the RHG lattice, and (b) an element of the stabilizer of an RHG graph state. By repeating the elementary cell, we obtain the RHG lattice illustrated in Fig.~\ref{fig:intro}. For a fixed vertex (labeled $v$), there are four adjacent vertices (labeled $u_1,u_2,u_3,u_4$). The RHG graph state is stabilized by $S_{v}=X_{v}\otimes\bigotimes_{k\in\{1,2,3,4\}}Z_{u_k}$.}
    \label{fig:stabilizer}
\end{figure}
%%%%%%%%%%%%%%%%%%%%%%%%%%%%

\textit{Preliminaries and notation.}---
Let $G_N=(V_N,E_N)$ be a graph in a family of undirected graphs on $|V_N|=N$ vertices. The $N$-qubit graph state $\ket{G_N}$ is defined by preparing a qubit in the $\ket{+}\propto\Ket{0}+\Ket{1}$ state for each vertex $v\in V_N$, and performing a controlled $Z$ gate between each pair of vertices $u,v\in V_N$ connected by an edge $\{u,v\}\in E_N$~\cite{H13}.

We write single-qubit Pauli operators as $X$, $Y$, and $Z$, and the identity operator as $\mathbbm{1}$. The support of a multiqubit Pauli operator is defined as the subset of qubits on which it acts as a non-identity Pauli operator, and its weight is the cardinality of its support.
The graph state is a stabilizer state with stabilizer (i.e., maximal abelian subgroup of the multiqubit Pauli group that leaves the state invariant) generated by the set of $N$-qubit Pauli operators~\cite{H13} 
\begin{equation}
\Big\{S_v \coloneqq X_v\otimes\bigotimes_{u\in\mathrm{Nbd}(v)}Z_u: v\in V_N\Big\},
\end{equation}
where $\mathbbm{1}$ is omitted, the subscripts of $X_v$ and $Z_u$ indicate they act on the qubits representing vertices $v$ and $u$ respectively, and the neighbourhood
\begin{equation}
       \mathrm{Nbd}(v) \coloneqq \{u\in V_N :\{u,v\}\in E_N\}
\end{equation}
 of $v$ is the set of vertices adjacent to $v$. Thus, $S_v$ acts as $X$ on the qubit labeled by $v$, as $Z$ on the vertices adjacent to $v$, and identity everywhere else. Notice that for a $D$-regular graph, each $S_v$ has weight $D+1$. 
 
 The Raussendorf-Harrington-Goyal (RHG) lattice~\cite{RAUSSENDORF20062242} is a $3$D lattice generated by the unit cell shown in Fig.~\ref{fig:stabilizer}. We will call graph states associated with connected subgraphs of this lattice \textit{RHG graph states}. The degree of all vertices in these graphs is $4$, except for those on boundaries. RHG graph states constitute a well-known family of resource states for fault-tolerant MBQC~\cite{RAUSSENDORF20062242,PhysRevLett.98.190504,Raussendorf_2007}. 
\\

\textit{Error model.}--- It is a common situation in implementing quantum computation that a source may be claimed to prepare some fixed desired state in the ideal case. However, the source may be faulty, and the state it prepares may suffer from errors which we can capture with a suitable error model. As in Ref.~\cite{RAUSSENDORF20062242}, we consider an error model described by IID depolarizing channels acting on each qubit as
\begin{equation}
\label{eq:error}
    \mathcal{N}_{p}(\sigma)=(1-p)\sigma+\frac{p}{3}(X\sigma X+Y\sigma Y+Z\sigma Z).
\end{equation}
The probability $p$ of a single qubit having an error is called the physical error rate.
We call an $X$ error a bit flip and a $Z$ error a phase flip.
The error $Y\propto XZ$ can be considered to be a combination of simultaneous bit- and phase-flip errors.
Following the standard MBQC setting, we may perform arbitrary single-qubit measurements on any qubit of the state from the source and additional classical computation using the measurement outcomes.
As in the threshold analysis in Ref.~\cite{RAUSSENDORF20062242}, we assume that measurements do not suffer from errors; however, we remark that our test protocol only uses measurements in the $X$ and $Z$ bases, and bit- and phase-flip errors in the measurement outcomes correspond to the respective Pauli errors in the state.

This error model is simple but also of practical importance in the threshold analysis of state-of-the-art fault-tolerant MBQC protocols using the Gottesman-Kitaev-Preskill (GKP) code~\cite{PhysRevA.64.012310,PhysRevA.102.032408} on photonic systems.
In photonic MBQC, each qubit of the graph state can be encoded in a continuous-variable (CV) mode of light by the GKP code, and single-qubit Pauli measurements are implementable by means of the well-established technology of homodyne detection~\cite{yamasaki2020polylogoverhead,PhysRevX.8.021054,Bourassa2021blueprintscalable,PhysRevA.64.012310,PhysRevA.102.032408,PhysRevLett.112.120504,PhysRevLett.123.200502,PhysRevResearch.2.023270}.
On this platform of photonic MBQC, physical errors on the CV systems indeed reduce to the IID Pauli errors on the GKP code~\cite{PhysRevLett.112.120504}.

It is known that if the physical error rate $p$ in~\eqref{eq:error} is below a threshold $p_\mathrm{th}\approx 1.4\times 10^{-2}$, a fault-tolerant MBQC protocol using RHG graph states can arbitrarily suppress the resulting logical error rate in simulating universal quantum computation~\cite{RAUSSENDORF20062242}.
On the other hand, if the physical error rate is just around the threshold $p_\mathrm{th}$, the fault-tolerant protocol may incur an excessive overhead~\cite{Preskill2018quantumcomputingin,PRXQuantum.1.020312,Gidney2021howtofactorbit}.
Thus in practice, we need a sufficiently better physical error rate $p_\mathrm{goal}(<p_\mathrm{th})$ --- and indeed, this is considered to be an important technological goal in experiments of realizing fault-tolerant quantum computation, beyond surpassing the threshold.\\

\textit{Task of testing graph states.}---
We will now formulate the task of testing graph states.
Consider a source of quantum states that is claimed to prepare an $N$-qubit graph state $\ket{G_N}$. The state preparation may be afflicted by the errors, and this is what we wish to test. Suppose that we have access to a single copy of the $N$-qubit state $\rho$ emitted by the source.
If no error occurs, $\rho$ is the pure state $\ket{G_N}$; on the other hand, errors that do occur will leave a signature on the state, which may be used to deduce properties of $\rho$ suffering from the noise channel~\eqref{eq:error} at physical error rate $p$.
We perform single-qubit measurements on $\rho$ and classical computation to conduct the testing task.

Of particular interest is the following task: given a significance level $\delta>0$, and thresholds $0<p_{\mathrm{goal}}<p_{\mathrm{th}}<1$, with probability at least $1-\delta$,
\begin{enumerate}
    \item \textsc{reject} if $p>p_\mathrm{th}$;
    \item \textsc{accept} if $p<p_\mathrm{goal}$.
\end{enumerate}
The threshold $p_\mathrm{th}$ is determined by the graph state and the fault-tolerant MBQC protocol to be used.
We can fix the significance level as desired, e.g., $\delta=\nicefrac{1}{3}$. As motivated previously, this task is tailored such that states on which we accept have a physical error rate that is small enough to be able to make fault-tolerant MBQC feasible.\\

\textit{Test protocol.}---
For any $p_\mathrm{th}$ and $\delta$, we now proceed to present a test protocol that efficiently accomplishes the above task for the input graph state $\Ket{G_N}$ of any size $N$.
Our test uses single-qubit measurements on $N_\mathrm{test}$ qubits of the $N$-qubit input state, and checks whether the $N_\mathrm{test}$-bit string of measurement outcomes satisfies a certain success condition by a simple classical computation. Remarkably, we will show that this is possible with a choice of $N_\mathrm{test}$ that is independent of the size $N$ of the input state, for values of the lower threshold $p_\mathrm{goal}\in(0,p_\mathrm{th})$ that are sufficiently gapped from $p_\mathrm{th}$ by an amount that depends only on $\delta$.

We first choose a subset of vertices $V_\mathrm{test}\subset V_N$ of size $|V_\mathrm{test}|=N_\mathrm{test}$ that have the same degree $D$ and are at a distance of at least three from each other.
The choice of $V_\mathrm{test}$ can be arbitrary as long as it satisfies these same-degree and distance-three constraints.
Note that for any family of bounded-degree periodic graphs $G_N$, we can always satisfy these constraints when $N$ is sufficiently large.
Given a single copy of the state $\rho$ from the source, measure the $N_\mathrm{test}$ qubits corresponding to vertices in $V_\mathrm{test}$ in the $X$ basis $\{\Ket{\pm}\}$, and measure all the qubits adjacent to these $N_\mathrm{test}$ qubits in the $Z$ basis $\{\Ket{0},\Ket{1}\}$.
The total number of qubits to be measured is
\begin{equation}
\label{eq:required_number_of_qubits}
    (D+1)N_\mathrm{test}.
\end{equation}
From the outcomes, calculate the \textit{parity} of each of the $N_\mathrm{test}$ stabilizer generators $S_v=X_v\otimes\bigotimes_{u\in \mathrm{Nbd}(v)}Z_u$ for $v\in V_\mathrm{test}$ in the following way:
\begin{enumerate}
    \item from the outcome $\{\Ket{\pm}\}$ of the $X$-basis measurement, set $b_v=\pm 1$;
    \item from the outcome $\{\Ket{j}:j=0,1\}$ of each $Z$-basis measurement, set $b_u={(-1)}^{j}$; and
    \item calculate the parity of $S_v$ as the product \begin{equation}
    b_v\times\prod_{u\in \mathrm{Nbd}(v)}b_u.
    \end{equation}
\end{enumerate}
Finally, the protocol \textsc{accepts} if the parity of $S_v$ is $+1$ for \textit{all} the $N_\mathrm{test}$ stabilizer generators indexed by $v\in V_\mathrm{test}$; otherwise it \textsc{rejects}. We analyze the correctness and resource requirement of this protocol below.

We remark that our test protocol is applicable to graph states represented by any family of bounded-degree periodic graphs $G_N$, which is general enough to include all the known graph states used for fault-tolerant MBQC~\cite{RAUSSENDORF20062242,PhysRevLett.98.190504,Raussendorf_2007,PhysRevLett.117.070501,nickerson2018measurement,Newman2020generatingfault}; for concreteness, we consider RHG graph states in the following analysis.\\

\textit{Correctness and Complexity.}---Given $p_\mathrm{th}\in(0,1)$ and $\delta>0$, we first provide a general prescription to obtain $N_\mathrm{test}$ and $p_\mathrm{goal}$, which will turn out to be constants that are independent of $N$. Subsequently, we will demonstrate a concrete realization of this prescription for the RHG graph states.

If no error has occurred, the parity of $S_v$ is always $+1$ since $S_v$ is in the stabilizer of $\Ket{G_N}$.
Nontrivial Pauli errors on some of the $(D+1)$ qubits in the support of $S_v$ may change the parity of $S_v$ into $-1$.  
In particular, let $E$ be a multiqubit Pauli operator acting nontrivially on a subset of the support of $S_v$.
If $E$ and $S_v$ anti-commute, i.e., $E S_v=-S_v E$, then the parity of $S_v$ becomes $-1$.
The probability of the event of having the error $E$ is given by
\begin{equation}
\label{eq:weight}
    p(E)\coloneqq\left(\frac{p}{3}\right)^w(1-p)^{1-w},
\end{equation}
where $w$ is the weight of $E$.
Then, the probability of flipping the parity of $S_v$ for each $v\in V_\mathrm{test}$ is
\begin{align}
\label{eq:p_flip_D}
    p_{\mathrm{flip}}(D)\coloneqq\sum_{E:E S_v=-S_v E}p(E),
\end{align}
where, for IID errors, $p_{\mathrm{flip}}(D)$ is independent of the identity of vertex $v$ and depends only on its degree. 
Since the distance between any pair of vertices $u,v\in V_\mathrm{test}$ is at least three, $S_u$ and $S_v$ have disjoint supports; that is, $p_{\mathrm{flip}}(D)$ is the same for $S_u$ and $S_v$. 

Next, we seek to bound $p_{\mathrm{flip}}(D)$ in terms of the physical error rate $p$, i.e.,
\begin{equation}
    \label{eq:general_bounds}
    l_D(p)\leqq p_{\mathrm{flip}}(D)\leqq u_D(p),
\end{equation}
where the lower bound $l_D(p)$ and the upper bound $u_D(p)$ are nondecreasing functions of $p$ which are completely determined by $D$.

In the case of high physical error rates $p>p_\mathrm{th}$, we want the protocol to \textsc{reject} with high probability, at least $1-\delta$.
Accordingly, \textit{at least one} of the parity values corresponding to the $N_\mathrm{test}$ stabilizer generators should be flipped, i.e.,
\begin{equation}
\label{eq:requirement_failure}
    1-(1-p_{\mathrm{flip}}(D))^{N_\mathrm{test}} \geqq 1-\delta.
\end{equation}
Since $\log (1-x) \leqq -x$ for $x<1$,~\eqref{eq:requirement_failure} is satisfied if
\begin{equation}
    N_\mathrm{test} \geqq \frac{\log(\nicefrac{1}{\delta})}{p_{\mathrm{flip}}(D)} \geqq \frac{\log(\delta)}{\log (1-p_{\mathrm{flip}}(D))},
\end{equation}
where the base of $\log$ is $\mathrm{e}$ unless specified otherwise.
Rewriting in terms of the lower bound  $p_\mathrm{flip}(D)\geqq l_D(p_\mathrm{th})$ from~\eqref{eq:general_bounds}, we have~\eqref{eq:requirement_failure} if
\begin{equation}
\label{eq:lower_N_0}
    N_\mathrm{test} \geqq \frac{\log(\nicefrac{1}{\delta})}{l_D(p_\mathrm{th})}.
\end{equation}
This can be satisfied by the choice
\begin{equation}
\label{eq:N_0_general}
    N_\mathrm{test}=\left\lceil\frac{\log(\nicefrac{1}{\delta})}{l_D(p_\mathrm{th})}\right\rceil, 
\end{equation}
where $\lceil x\rceil$ is smallest integer larger than $x$.

In the case of low error rates $p<p_\mathrm{goal}$, we require the protocol to \textsc{accept} with high probability, at least $1-\delta$.
Correspondingly, \textit{none} of the $N_\mathrm{test}$ parity values should be flipped, i.e.,
\begin{align}
\label{eq:requirement_success}
    (1-p_\mathrm{flip}(D))^{N_\mathrm{test}} \geqq 1-\delta.
\end{align}
This requirement is satisfied if
\begin{align}
    N_\mathrm{test}\log\left(1-p_\mathrm{flip}(D)\right) &\geqq -\delta \geqq \log\left(1-\delta\right).
\end{align}
Thus, since $-x\leqq(1-x)\log(1-x)$ for $x<1$,~\eqref{eq:requirement_success} is satisfied if
\begin{align}
    N_\mathrm{test}\leqq\frac{1-p_\mathrm{flip}(D)}{p_\mathrm{flip}(D)}\times\delta.
\end{align}
Using the upper bound $p_\mathrm{flip}(D)\leqq u_D(p_\mathrm{goal})$ from~\eqref{eq:general_bounds},
we see that~\eqref{eq:requirement_success} will hold if $p_\mathrm{goal}$ is sufficiently small, as given by
\begin{align}
\label{eq:upper_N_0}
    N_\mathrm{test} \leqq \frac{1-u_D(p_\mathrm{goal})}{u_D(p_\mathrm{goal})}\times \delta.
\end{align}
In order to be able to pick a value of $N_\mathrm{test}$ that satisfies both the lower bound~\eqref{eq:lower_N_0} and the upper bound~\eqref{eq:upper_N_0}, $p_{\mathrm{goal}}$ should satisfy
\begin{align}
\label{eq:p_goal_general}
    \frac{u_D(p_{\mathrm{goal}})}{1-u_D(p_{\mathrm{goal}})}\leqq\frac{\delta}{\log(\nicefrac{1}{\delta})}\times l_D(p_\mathrm{th}).
\end{align}
For specific realizations of $l_D$ and $u_D$, solving this inequality with respect to $p_\mathrm{goal}$ gives us a bound on the gap $p_\mathrm{th}-p_\mathrm{goal}$ required for our test protocol. 

%%%%%%%%%%%%%%%%%%%%%%%%%%%%%%%%%%%%%%
\begin{table}[t]
    \centering
    \begin{tabular}{l|cccccc}
    \hline
    weight $w$ of $E$ on support of $S_v$ & 0 & 1 & 2 & 3 & 4 & 5 \\
    \hline
    number of commuting $E$ & 1 & 5 & 50 & 130 & 205 & 121 \\
    number of anti-commuting $E$ & 0 & 10 & 40 & 140 & 200 & 122\\ 
    \hline
    \end{tabular}
    \caption{Numbers of Pauli errors $E$ that commute and anti-commute with a stabilizer element $S_v=X_{v}\otimes\bigotimes_{k\in\{1,2,3,4\}}Z_{u_k}$ at a degree-$4$ vertex $v$ of RHG graph states (Fig.~\ref{fig:stabilizer}). The table counts the numbers of $5$-qubit operators in $\{P_{v}\otimes\bigotimes_{k\in\{1,2,3,4\}}P_{u_k}:P_{v},P_{u_k}\in\{X,Y,Z,\mathbbm{1}\}\}$ acting on the support of $S_v$, i.e., those relevant to the commutativity. For a Pauli error $E$ of weight $w$, the probability of the error $E$ occurring is given by~\eqref{eq:weight}. The number of commuting errors $ES_v=S_v E$ and anti-commuting errors $ES_v=-S_v E$ together yields the probability $p_\mathrm{flip}$ of flipping the parity of $S_v$ as shown in~\eqref{eq:pflip_D_4}.}
    \label{tab:number_of_events}
\end{table}
%%%%%%%%%%%%%%%%%%%%%%%%%%%%%%%%%%%%%%

For example, for the RHG graph states with $D=4$, the numbers of errors $E$ that commute and anti-commute with $S_v$ are summarized in Table~\ref{tab:number_of_events}\@.
Thus, due to~\eqref{eq:p_flip_D}, $p_\mathrm{flip}(4)$ is given by
\begin{align}
\label{eq:pflip_D_4}
    p_\mathrm{flip}(4)&=10{\left(\frac{p}{3}\right)}(1-p)^4+40{\left(\frac{p}{3}\right)}^2(1-p)^3+\nonumber\\
    &140{\left(\frac{p}{3}\right)}^3(1-p)^2+200{\left(\frac{p}{3}\right)}^4(1-p)+122{\left(\frac{p}{3}\right)}^5.
\end{align}
It holds that (assuming $p_{\mathrm{th}}<\nicefrac{3}{8}$)
\begin{equation}
    \label{eq:pflip-bounds}
    \frac{10}{3}p-\frac{80}{9}p^2\leqq p_\mathrm{flip}(4)\leqq\frac{10}{3}p;
\end{equation}
accordingly, we take
\begin{align}
    l_4(p)&=\frac{10}{3}p-\frac{80}{9}p^2,\\
    u_4(p)&=\frac{10}{3}p.
\end{align}
Then, following~\eqref{eq:N_0_general}, we give $N_\mathrm{test}$ by
\begin{equation}
\label{eq:N_0_D_4}
    N_\mathrm{test}=\left\lceil\frac{\log(\nicefrac{1}{\delta})}{\frac{10}{3}p_\mathrm{th}-\frac{80}{9}p_\mathrm{th}^2}\right\rceil.
\end{equation}
As for $p_\mathrm{goal}$,~\eqref{eq:p_goal_general} yields the condition
\begin{equation}
    \frac{\frac{10}{3}p_\mathrm{goal}}{1-\frac{10}{3}p_\mathrm{goal}}\leqq\frac{\delta}{\log(\nicefrac{1}{\delta})}\left(\frac{10}{3}p_\mathrm{th}-\frac{80}{9}p_\mathrm{th}^2\right);
\end{equation}
i.e., it suffices to take
\begin{equation}
\label{eq:p_goal_D_4}
    p_\mathrm{goal}=\frac{3}{10}\times\frac{\frac{\delta}{\log(\nicefrac{1}{\delta})}\left(\frac{10}{3}p_\mathrm{th}-\frac{80}{9}p_\mathrm{th}^2\right)}{1+\frac{\delta}{\log(\nicefrac{1}{\delta})}\left(\frac{10}{3}p_\mathrm{th}-\frac{80}{9}p_\mathrm{th}^2\right)}.
\end{equation}
Consequently, for small $p_\mathrm{th}$, we typically have 
\begin{equation}
    p_{\mathrm{goal}} = \frac{\delta}{\log(\nicefrac{1}{\delta})}\times p_{\mathrm{th}}+o(p_{\mathrm{th}}).
\end{equation}
The right-hand side indicates that, for a conventional choice of $\delta=\nicefrac{1}{3}$, we can reliably test the case of $p>p_{\mathrm{th}}$ against the case of $p<p_{\mathrm{goal}}$ as long as \begin{equation}
\label{eq:gap}
    p_{\mathrm{goal}}\lessapprox 0.3 p_{\mathrm{th}},\quad\text{i.e.,}\quad\ p_{\mathrm{th}}-p_{\mathrm{goal}}\gtrapprox 0.7p_{\mathrm{th}},
\end{equation} 
which is to say we can relatively easily distinguish whether the state prepared by the source is `bad' or `very good'.

Remarkably, $N_\mathrm{test}$ can be as small as $O(1)$ independently of $N$, and after measuring $(D+1)N_\mathrm{test}$ qubits in~\eqref{eq:required_number_of_qubits}, we can use the remaining $(N-(D+1)N_\mathrm{test})$-qubit graph state for fault-tolerant MBQC\@.
For example, in the case of $\delta=\nicefrac{1}{3}$, $D=4$, and $p_\mathrm{th}=1.4\times 10^{-2}$ as in the RHG lattice~\cite{RAUSSENDORF20062242},
due to~\eqref{eq:N_0_D_4} and~\eqref{eq:p_goal_D_4},
it suffices to choose
\begin{align}
    N_\mathrm{test}&=25,\\
    p_\mathrm{goal}&=4.0\times 10^{-3};
\end{align}
then, the test requires only $125$ qubits.
These values are expected to be within the reach of near-term quantum devices~\cite{Preskill2018quantumcomputingin}.\\

\textit{Discussion and outlook.}--- 
One na\"ive way to address the task of discriminating high and low physical error rates $p$ is to completely \textit{learn} the underlying noise channel, e.g.,~\eqref{eq:error}.
While this task can in principle be done by standard methods for channel estimation and tomography~\cite{Flammia2020,Harper2020}, these techniques require the ability to study the input-output behaviour of the noise channel for many different choices of input, making them unfavourable for the photonic-MBQC platform using the fixed input graph state.
Furthermore, these techniques require a quadratically large complexity $O(\nicefrac{1}{\epsilon^2})$ to learn a single-qubit channel with precision $\epsilon$ to which the channel is estimated.
By contrast, in the case of~\eqref{eq:gap} with the fixed $\delta$, our protocol can discriminate $p>p_\mathrm{th}$ and $p\lessapprox 0.3 p_\mathrm{th}$ only within a linear complexity $N_\mathrm{test}=O(\nicefrac{1}{p_\mathrm{th}})$, as shown in~\eqref{eq:N_0_D_4}.
Therefore, our results achieve a quadratic improvement over the channel estimation with precision $\epsilon\approx p_\mathrm{th}$ and complexity $O(\nicefrac{1}{p_\mathrm{th}^2})$ for this discrimination. 

Our analysis has demonstrated the feasibility of the test for IID Pauli errors on the graph states; however, we remark that the formulation and the analysis are potentially applicable to more general error models such as biased noise, circuit-level errors, adversarial correlated errors, and CV Gaussian errors in using the GKP code on photonic systems, and to other classes of states such as hypergraph states, magic states, and states in the codespace of the stabilizer code.\\

\textit{Conclusion.}---
We have developed a framework for testing preparation of graph states for fault-tolerant MBQC, in the setting of IID Pauli errors that are conventional in fault-tolerant quantum computation.
In contrast to past work that focuses on fidelity estimation for verifying that states prepared by a source are close to the required resource state, our focus is on distinguishing states with high and low values of the physical error rate. This approach is well motivated by the fact that ultimately, for fault-tolerant MBQC to be realized efficiently, it is essential that the physical error rate be smaller than and bounded away from the threshold.
For any graph state represented by a bounded-degree periodic graph, i.e, that used for fault-tolerant MBQC,
our test protocol only uses a single copy of the graph state and performs single-qubit measurements on a constant number of qubits, so that the rest of the tested and accepted graph state can be used subsequently for fault-tolerant MBQC\@.
This protocol only requires a constant runtime regardless of the number of qubits of the entire graph state.
Thus, our protocol leads to a significant advantage over the existing verification protocols based on fidelity estimation that have required many copies of the entire graph states and hence required an excessive number of qubits and runtime.
These results open a novel route to a practically feasible framework to benchmark the preparation of large-scale entangled states.\\

\begin{acknowledgments}
HY is supported by JSPS Overseas Research Fellowships and JST PRESTO Grant Number JPMJPR201A\@. SS is supported by Tom Gur's UKRI Future
Leaders Fellowship MR/S031545/1.
\end{acknowledgments}

\bibliography{sample}

\end{document}